\documentclass[a4paper,11pt]{article}
\usepackage{jheppub} 
\usepackage{bbold}
\usepackage[T1]{fontenc} 
\newcommand{\Trh}{T_\text{rh}}
\newcommand{\YBL}{Y_{B-L}}
\newcommand{\gs}{g_\star}
\newcommand{\gss}{g_{\star s}}
\newcommand{\phicmb}{\phi_{\text{CMB}}}
\newcommand{\ncmb}{N_{\text{CMB}}}
\newcommand{\phiend}{\phi_{\text{end}}}
\newcommand{\nbl}{n_{B-L}}

\title{Parameter Space of Leptogenesis in Polynomial Inflation}
\author[a]{Manuel Drees}
\author[b]{and Yong Xu}
\affiliation[a]{\it Bethe Center for Theoretical Physics and Physikalisches
	Institut, Universit\"at Bonn\\ Nussallee~12, 53115 Bonn, Germany}
\affiliation[b]{\it PRISMA$^{+}$ Cluster of Excellence and Mainz Institute for
	Theoretical
	Physics, Johannes Gutenberg University, 55099 Mainz, Germany}
\emailAdd{drees@th.physik.uni-bonn.de}
\emailAdd{yonxu@uni-mainz.de}
\abstract{Polynomial inflation is a very simple and well motivated
	scenario. A potential with a concave ``almost'' saddle point at
	field value $\phi = \phi_0$ fits well the cosmic microwave
	background (CMB) data and makes testable predictions for the running
	of the spectral index and the tensor to scalar ratio. In this work we analyze leptogenesis in the polynomial inflation framework. We delineate the allowed parameter space giving rise to the correct baryon asymmetry as well
	as being consistent with data on neutrino oscillations. To that end
	we consider two different reheating scenarios. $(i)$ If the inflaton
	decays into two bosons, the reheating temperature can be as high as
	$\Trh \sim 10^{14}$ GeV without spoiling the flatness of the
	potential, allowing vanilla $N_1$ thermal leptogenesis to work if
	$\Trh > M_1$ where $N_1$ is the lightest right--handed neutrino and
	$M_1$ its mass. Moreover, if the dominant decay of the inflaton is
	into Higgs bosons of the Standard Model, we find that rare
	three--body inflaton decays into a Higgs boson plus one light and
	one heavy neutrino allow leptogenesis even for $\Trh < M_1$ if the
	inflaton mass is of order $10^{12}$ GeV or higher; in the polynomial
	inflation scenario this requires $\phi_0 \gtrsim 2.5~M_P$. This
	novel mechanism of non--thermal leptogenesis is quite generic, since
	the coupling leading to the three--body final state is required in
	the type I see--saw mechanism. $(ii)$ If the inflaton decays into
	two fermions, the flatness of the potential implies a lower
	reheating temperature. In this case inflaton decay to two $N_1$
	still allows successful non--thermal leptogenesis if
	$\phi_0 \gtrsim 0.1~M_P$ and $\Trh \gtrsim 10^{6}$ GeV.}

\begin{document} 
	\begin{flushright}
		MITP-24-001\\
	February 2024
	\end{flushright}
	\maketitle
	\flushbottom
	\section{Introduction}
	
	Monomial inflationary scenarios have been ruled out by recent Planck
	and BICEP/Keck 2018 measurements \cite{Planck:2018jri,
		BICEP:2021xfz}. We have recently shown that a simple polynomial of
	degree four with a concave “almost'' saddle point at field value
	$\phi = \phi_0$ can nevertheless fit well the CMB data
	\cite{Drees:2021wgd, Drees:2022aea}. The model predicts the running of
	the spectral index to be $\mathcal{O}(10^{-3})$, which may be testable
	by the next generation of cosmic microwave background (CMB)
	experiments in combination with data on structure formation at smaller
	scales \cite{Munoz:2016owz}. The predicted tensor to scalar ratio $r$
	(in the large field scenario) can saturate the current bound.
	
	A full model of the very early Universe also has to describe
	reheating, i.e. transition to a thermal stage. We have considered two
	simple scenarios of perturbative reheating where the inflaton decays
	via renormalizable couplings, either to a pair of bosons [e.g. the
	standard model (SM) Higgs] or to vector--like fermions
	[e.g. right-handed-neutrinos (RHNs)]. We have obtained the full
	allowed parameter space of the model by $(i)$ considering the lower
	bound on $\Trh$ from Big Bang nucleosynthesis (BBN) as well as the
	stability of the inflaton potential against radiative corrections
	which leads to an upper bound on $\Trh$ for given $\phi_0$; and $(ii)$
	confronting the model with the latest CMB data, which leads to an
	upper bound on $\phi_0$ and hence on the inflaton mass. In combination
	we find that the inflection point has to satisfy
	$3\times 10^{-5} \lesssim \phi_0/M_P \lesssim 21.5$, which also
	allows us to derive the allowed range of reheating temperature. It can
	be as large as $\Trh \sim 10^{14}$ GeV for bosonic inflaton decays.
	
	Two further ingredients are required for a realistic cosmological
	model: it has to explain the existence of Dark Matter (DM), and it has
	to generate an excess of baryons over antibaryons. Recently it has
	been shown \cite{Bernal:2021qrl} that in polynomial inflation a rather
	large parameter space exists for either thermal or non--thermal DM
	production after inflation. In this work we show that the model can
	also produce the baryon asymmetry. To this end, we focus on the most
	simple ``vanilla leptogenesis'' in the framework of the type I seesaw
	mechanism. Here a lepton asymmetry is generated (solely) from the
	decay of the lightest right--handed neutrino $N_1$, and the model
	parameters are chosen such that the measured squared mass differences
	and mixing angles of the light neutrinos are reproduced. $N_1$ can be
	produced either thermally \cite{Fukugita:1986hr, Buchmuller:2004nz,
		Covi:1996wh, Fong:2012buy} or non--thermally (via inflaton decay)
	\cite{Lazarides:1990huy, Giudice:1999fb, Asaka:1999yd, Asaka:1999jb,
		Senoguz:2003hc, Hahn-Woernle:2008tsk, Antusch:2010mv, Croon:2019dfw}, depending on its mass and on the reheating
	temperature.
	
	In the framework of non--thermal leptogenesis, so far only the
	inflaton two body decay to RHN, $\phi \rightarrow N_1 N_1$, has been
	considered in the literature, see e.g. Refs.~\cite{Lazarides:1990huy,
		Giudice:1999fb, Asaka:1999yd, Asaka:1999jb, Senoguz:2003hc,
		Hahn-Woernle:2008tsk, Antusch:2010mv, Croon:2019dfw}. Here we point
	out that non--thermal leptogenesis is possible even if the dominant
	inflaton decay is into two SM Higgs bosons $H$; this is the simplest,
	and hence best motivated, bosonic inflaton decay mode. Type I seesaw
	requires the existence of $H N \nu$ couplings, so inflaton decays
	$\phi \to H \ell N_1$ will occur, albeit with suppressed rate, as
	long as $m_\phi > M_1$. In this work we will show that such three body
	decays can also give rise to successful non--thermal leptogenesis even
	without introducing any direct couplings between the inflaton and
	RHNs. We analyze this scenario again for polynomial inflation;
	however, the only relevant inflationary parameters are the reheat
	temperature and the inflaton mass, so this mechanism obviously can be
	applied to other models of inflation as well.
	
	The remainder of this paper is organized as follows. In
	Sec.~\ref{setup} we introduce the Lagrangian of our model. In
	Sec.~\ref{inflation} we briefly review polynomial inflation
	\cite{Drees:2021wgd, Drees:2022aea}. Reheating is reviewed in
	Sec.~\ref{reheating}, while the masses and mixing angles of the light
	neutrinos are discussed in Sec.~\ref{numass}, with focus on normal
	ordering (i.e. an effectively massless lightest neutrino). In
	Sec.~\ref{leptogenesis} we investigate both thermal and non--thermal
	channels for leptogenesis. We sum up our findings in
	Sec.~\ref{sum}. The calculation of the branching ratio for three--body
	inflaton decays is described in the Appendix.
	
	
	\section{The Model Setup}
	\label{setup}
	
	Applying Occam's razor \cite{Occam:1495nn}, we minimally extend the
	Standard Model (SM) with a scalar inflaton field $\phi$ and three
	right-handed Majorana neutrinos (RHNs) $N_I$ with $I=1,2,3$. Insisting
	that all non--gravitational interactions are renormalizable leads to
	the following action:
	\begin{equation} \label{eq:S}
	S = \int d^4x\, \sqrt{-g} \left( \mathcal{L}_{\text{Gravity}}
	+  \mathcal{L}_{\rm SM}  + \mathcal{L}_{\phi} + \mathcal{L}_{N}
	+ \mathcal{L}_{H\phi} \right)\,.
	\end{equation}
	Here $g$ is the determinant of the metric; we assume the latter to have the
	Friedmann--Lemaître--Robertson--Walker (FLRW) form, defined by
	$g_{\mu \nu} = \text{diag}(1,-a^2,-a^2,-a^2)$, with $a$ denoting the
	scale factor. Moreover, $H$ in eq.(\ref{eq:S}) denotes the SM Higgs
	doublet, and $\mathcal{L}_{\text{Gravity}}$ is the Einstein--Hilbert
	Lagrangian:
	\begin{equation} \label{eq:EH}
	\mathcal{L}_{\text{Gravity}}   = \frac{M_P^2}{2}\, R\,,
	\end{equation}
	with $R$ being the Ricci scalar and $M_P \simeq 2.4 \times 10^{18}$~GeV
	the reduced Planck mass.
	
	Furthermore, $\mathcal{L}_{\rm SM}$ in eq.(\ref{eq:S}) is the usual SM
	Lagrangian, while $\mathcal{L}_\phi$ describes the inflaton sector:
	\begin{equation}
	\mathcal{L}_\phi = \frac{1}{2} \partial_\mu \phi \partial^\mu \phi - V(\phi)\,.
	\end{equation}
	We consider the most general renormalizable inflaton potential
	$V(\phi)$:\footnote{A possible linear term can be absorbed by shifting the
		field $\phi$, and a possible additive constant should be tiny in order
		to produce an (almost) vanishing cosmological constant after inflation.}
	\begin{equation} \label{inflaton_potential}
	V(\phi)= b\, \phi^2 + c\, \phi^3 + d\, \phi^4.
	\end{equation}
	The SM and inflation sectors can couple via
	\begin{equation} \label{Vphih}
	\mathcal{L}_{H \phi} = -V(H,\phi) =  -\lambda_{12}\, \phi\, H^{\dagger} H
	- \frac{1}{2} \lambda_{22}\, \phi^2\, H^{\dagger} H  \,;
	\end{equation}
	in particular, the coupling $\lambda_{12}$ allows perturbative reheating
	via $\phi \rightarrow H H$ decays. 
	
	The final ingredient of the action (\ref{eq:S}) is the piece of the
	Lagrangian describing the RHNs, including possible interactions with the
	inflaton. It is given by:
	\begin{align} \label{LN}
	\mathcal{L}_N = i\, \overline{N_I}\, \gamma^\mu\, \partial_\mu N_I
	- \left( \frac {1} {2} M_I \overline{N^c_I} N_I + h.c.  \right)\,
	- \left( Y_{\alpha I}\, \bar{\ell}_\alpha \tilde{H} N_I + h.c. \right)
	- \left( \frac{y_I}{2}\, \phi \overline{N^c_I} N_I + h.c.  \right)\,.
	\end{align}
	We have worked in a basis with diagonal Majorana mass matrix for the
	$N_I$, the $M_I$ denoting its eigenvalues. Moreover,
	$\ell_\alpha,\; \alpha\in\{1,2,3\}$ are the lepton doublets,
	$\tilde{H} \equiv i\sigma_2 H^\star$, and $Y_{\alpha I}$ and $y_I$ are
	Yukawa couplings; the latter have been normalized such that $y_I$
	appears as $\phi N_I N_I$ vertex factor. This coupling might allow
	$\phi \rightarrow N_I N_I$ decays, a second possibility for
	perturbative reheating. We note that in full generality we should also
	allow off--diagonal $\phi N_I N_K$ couplings, since the coupling
	matrix to the inflaton and the $N_I$ mass matrix in general cannot be
	diagonalized simultaneously. However, in our application we will be
	interested in the lightest RHN in a hierarchical scenario,
	$M_1 \ll M_2,\, M_3$; here essentially only the $\phi N_1 N_1$
	coupling is relevant.
	
	Before closing this section, we would like to comment on the relation
	between the inflaton couplings to Higgs bosons and RHNs. Once one of
	these couplings is introduced, the other one will be generated by
	one--loop diagrams. Specifically, if $\lambda_{12} \neq 0$, a triangle
	diagram with two Higgs bosons and a lepton doublet in the loop will
	generate a finite coupling of the inflaton to two RHNs,
	$y^{\text{eff}}_I \sim \lambda_{12} \, M_I |Y|^2/(16 \pi^2 m_\phi^2)$,
	where $m_\phi^2$ comes from the denominators of the propagators and
	$M_I$ appears after utilizing the Dirac equation for the lepton
	momentum in the numerator. Conversely, if $y_I \neq 0$, a triangle
	diagram with two RHNs and one lepton doublet in the loop will generate
	a {\em divergent} inflaton couplings to two SM Higgs bosons,
	$\lambda_{12}^{\text{eff}} \propto y_I M_I |Y|^2$.
	
	The loop--induced $y^{\text{eff}}_I$ will generate a partial width for
	$\phi \rightarrow N N$ decays which is suppressed by a loop factor
	$|Y|^2 / (16 \pi^2)$ relative to the width for
	$\phi \rightarrow H N \ell$ decays that we will discuss in
	sec.\ref{sec:3body}; it will thus generally not be of great
	phenomenological importance. On the other hand, the fact that
	$y_I \neq 0$ generates a divergent contribution to $\lambda_{12}$
	means that a theory with nonvanishing coupling of the inflaton to RHNs
	but no inflaton coupling to SM Higgs bosons is, strictly speaking, not
	renormalizable. Choosing $\lambda_{12} \neq 0$ but $y_I = 0$ is
	therefore in some sense ``more minimal'', since it needs fewer
	fundamental parameters in the Lagrangian. However, the fact that a
	finite $y_I^{\text{eff}}$ will be generated means that this ``more
	minimal'' set--up cannot be ensured by a (conserved) symmetry.
	
	\section{Inflation}
	\label{inflation}
	
	In this section we review the polynomial inflation formalism presented
	recently in Refs.~\cite{Drees:2021wgd, Drees:2022aea}. The potential in
	eq.~\eqref{inflaton_potential} features an inflection point at
	$\phi = \phi_0 = -\frac{3}{8} \frac{c}{d}$ if
	$b = \frac{9}{32} \frac{c^2}{d}$. However, in order to match the CMB
	data the potential has to have a finite (positive) slope rather than
	an exact inflection point (where both the first and second derivatives
	of the potential vanish). Moreover, a potential with a concave shape
	is favored by the Planck 2018 data \cite{Planck:2018vyg}. We thus
	reparametrize the potential as
	\begin{align}  \label{inflaton_potential2}
	V(\phi) &=   d \left[ \phi^4 + \frac {c} {d} ( 1 - \beta ) \phi^3
	+  \frac {9} {32} \left( \frac {c} {d} \right)^2 \phi^2 \right]
	\nonumber \\
	& =d \left[ \phi^4 + A ( 1 - \beta ) \phi^3 + \frac {9} {32} A^2\phi^2\right]\,,
	\end{align}
	where $A\equiv \frac{c}{d} \equiv -\frac{8}{3} \phi_0$ determines the
	location of the flat region of the potential and $\beta$ is introduced
	in order to produce a finite slope near $\phi_0$; i.e. $\beta=0$ leads
	to a true inflection point at $\phi_0$. Successful inflation results
	for $0<\beta \ll 1$.
	
	We define the traditional potential slow roll (SR) parameters
	\cite{Lyth:2009zz}:
	\begin{align}\label{srparameters}
	\epsilon_V \equiv \frac {1} {2} \left( \frac{V^{\prime}}{V} \right)^2\,;
	\ \  \eta_V \equiv \frac {V^{\prime \prime}} {V}\,;\ \
	\xi_V^2 \equiv \frac {V^{\prime} V^{\prime \prime \prime}} {V^2}\,.
	\end{align}
	They do not depend on the overall normalization of the potential $d$,
	and must be small during inflation, i.e. $\epsilon_V$, $|\eta_V|$ and
	$|\xi_V^2| \ll 1$.
	
	We define the end of inflation at a field value $\phiend$ such that
	$\epsilon_V(\phiend)=1$. The total number of e--folds between the time
	when the CMB pivot scale $k_{\star} = 0.05\ \rm{Mpc}^{-1}$ first
	crossed out the horizon till end of inflation can be computed analytically
	in this model \cite{Drees:2022aea}:
	\begin{align} \label{ncmb}
	\ncmb &= \int^{\phicmb}_{\phiend} \frac{1}{\sqrt{2 \epsilon_V}} d\phi
	\nonumber \\
	&\simeq \frac{1}{24} \left \{ 3\phi^2 - 4\phi \phi_0 + 15\phi_0^2
	-\phi_0^2 \sqrt{ \frac{2}{\beta} }
	\arctan\left( \frac{\phi_0 -\phi }{\sqrt{2\beta}\phi_0} \right)
	- \phi_0^2 \ln\left[(\phi_0 - \phi)^2 \right] \right \}
	\Bigg \vert^{\phi_{\text{CMB}}}_{\phiend}\,.
	\end{align}
	Here $\phicmb$ is the field value when $k_\star$ first crossed out of
	the horizon. The predictions for the normalization
	$\mathcal{P}_{\zeta}$ of the power spectrum, its spectral index $n_s$,
	the running $\alpha$ of the spectral index, and the tensor to scalar
	ratio $r$ during SR inflation are given by \cite{Lyth:2009zz}
	\begin{align}
	\mathcal{P}_\zeta &= \frac{V}{24\pi^2\epsilon_V}\,;\\
	n_s &= 1 - 6 \epsilon_V + 2 \eta_V\,;\\
	\alpha &= 16 \epsilon_V \eta_V - 24 \epsilon_V^2 - 2 \xi_V^2\,;\\
	r &= 16 \epsilon_V\,.
	\end{align}
	The recent Planck 2018 measurements plus results on baryonic acoustic
	oscillations (BAO) at the pivot scale
	$k_{\star} = 0.05\ \rm{Mpc}^{-1}$ imply \cite{Planck:2018vyg}:
	\begin{equation}  \label{planck2018}
	\mathcal{P}_{\zeta} = (2.1 \pm 0.1) \cdot 10^{-9}\,; \
	n_s =  0.9659  \pm 0.0040\,; \  \alpha = -0.0041 \pm 0.0067\,.
	\end{equation} 
	The most stringent constraint on $r$ comes from BICEP/Keck 2018 (together with
	Planck data) \cite{BICEP:2021xfz}:
	\begin{equation}  \label{BK2018}
	r_{0.05} < 0.035\,\quad \text{at 95\% \text{C.L.}}\,.
	\end{equation} 
	%

	\section{Reheating}
	\label{reheating}
	
	In this section, we focus on two reheating scenarios, where the
	inflaton primarily decays either into fermions (specifically, into
	RHNs) or into Higgs bosons. The decay rate for inflaton decays to a
	pair of $N_1$ is given by
	\begin{equation} \label{eq:Gam_NN}
	\Gamma_{\phi \to \bar N_1 N_1}  = \frac { y_1^2\, m_\phi } {16\pi} \left(
	1 - \frac {4M_1^2} {m_\phi^2} \right)^{3/2}\,;
	\end{equation}
	for the bosonic channel one has
	\begin{equation} \label{eq:Gam_HH}
	\Gamma_{\phi \to H^{\dagger} H}  = \frac {\lambda_{12}^2} {8\pi\, m_\phi}
	\sqrt{1 - \frac {4m_H^2} {m_\phi^2}} \,.
	\end{equation}
	
	The reheating temperature $\Trh$, i.e. the highest temperature of the
	radiation dominated era, can be defined via
	$H(\Trh) = \frac23 \Gamma_\phi$, where $H(\Trh)$ denotes the Hubble
	parameter at $T = \Trh$. This gives
	\begin{equation}\label{trh}
	\Trh = \sqrt { \frac {2} {\pi} } \left( \frac {10} {\gs} \right)^{1/4}
	\sqrt {M_P\, \Gamma_\phi}\,,
	\end{equation}
	where $\gs(T)$ denotes the number of relativistic degrees of freedom
	contributing to the SM energy density $\rho_R$; in the SM,
	$\gs = 106.75$ for temperatures much higher than the electroweak
	scale. In order to allow successful BBN, the reheating temperature has
	to satisfy $\Trh \gtrsim 4~\text{MeV}$ \cite{Kawasaki:2000en,
		Hannestad:2004px}. On the other hand, the inflaton potential at $\phi_0$
	will be destabilized by one--loop corrections if the inflaton has too
	large couplings. Radiative stability implies~\cite{Drees:2021wgd}:
	\begin{equation} \label{eq:l12bound}
	\left| \left( \frac{ \lambda_{12}} {\phi_0} \right)^2 \ln \left( \frac
	{\lambda_{12}} {\phi_0} \right) - \left( \frac {\lambda_{12}} {\phi_0}
	\right)^2 \right| < 64\pi^2 d\, \beta \simeq 6.1 \times 10^{-19}
	\left( \frac {\phi_0} {M_P} \right)^6\,,
	\end{equation}
	and
	\begin{equation} \label{eq:y1bound}
	\left| y_1^4  - 3 y_1^4 \ln \left( y_1^2 \right) \right|
	< 128 \pi^2 d\, \beta \simeq 1.2 \times 10^{-18}
	\left( \frac {\phi_0}{M_P} \right)^6 \,.
	\end{equation}
	The approximate equalities hold for $\phi_0 \lesssim 3M_P$ where
	fixing $\phi_0$ essentially determines $\beta$ and $d$.
	
	Altogether, $\phi_0$ can lie in the rather wide range
	\begin{equation} \label{eq:phi0_range}
	3\times 10^{-5}\, M_P \lesssim \phi_0 \lesssim 21.5\, M_P\,.
	\end{equation}
	The lower bound comes from the combination of the lower bound
	$\Trh \gtrsim 4~\text{MeV}$ with the upper bounds of
	eq.\eqref{eq:l12bound} and eq.\eqref{eq:y1bound} \cite{Drees:2021wgd};
	note that the latter immediately imply upper bounds on $\Trh$ via
	eq.\eqref{trh}. The upper bound on $\phi_0$ in (\ref{eq:phi0_range})
	is essentially due to the upper bound (\ref{BK2018}) on $r$
	\cite{Drees:2022aea}.

	\section{Neutrino Masses}
	\label{numass}
	
	A simple method to derive the see--saw expression for the mass matrix
	of the light neutrinos is to integrate out the RHNs via their
	Euler--Lagrange equation,
	$\partial_\mu \frac{ \partial \mathcal{L}} {\partial (\partial_\mu
		\bar{N}_I)} - \frac{\partial \mathcal{L}}{\partial \bar{N}_I} = 0$.
	Solving for $N_I$ and plugging the result back into eq.\eqref{LN}, one
	finds (after electroweak symmetry breaking) a Majorana mass term for
	the light active neutrinos $\nu_\alpha$:
	\begin{equation} 
	\mathcal{L} \supset \bar{\nu}_{\alpha} \left( -2 \frac {v^2
		Y_{\alpha I} Y^T_{\alpha I} } {M_I} \right) \nu_\alpha^{c}\,,
	\end{equation}
	where $v=174$~GeV is the vacuum expectation value (vev) of the Higgs
	field. The mass matrix of the light neutrinos is therefore
	\begin{equation} \label{mnu1}
	\widetilde{m}_\nu = - v^2 Y M_I^{-1} Y^{T}\,,
	\end{equation}
	where $Y$ and $M_I$ are matrices. $\widetilde{m}_\nu$ can be
	diagonalized with the help of the Pontecorvo--Maki--Nakagawa--Sakata
	(PMNS) matrix\footnote{We work in the basis for the lepton doublets
		where the charged lepton mass matrix is diagonal.} $U$:
	\begin{equation} \label{eq:PMNS}
	U^T \widetilde{m}_\nu U = \text{diag}(m_1,m_2,m_3) \equiv m_\nu^{\rm diag}\,.
	\end{equation}
	In the following we assume the normal hierarchy for the light
	neutrinos where $m_1<m_2< m_3$, and consider the current best fit
	\cite{Esteban:2018azc}:
	\begin{align} \label{eq:numass2}
	m_1 &= 0\,; \nonumber \\
	m_2 &= \sqrt{ \Delta m_{\odot}^2} = ( 8.6 \pm 0.1 ) \times 10^{-3}\,
	\mathrm{eV}\,, \nonumber \\
	m_3 &= \sqrt{ \Delta m_{\mathrm{atm}}^2} = (5.02 \pm 0.03 )
	\times 10^{-2}\, \mathrm{eV}\,.
	\end{align}

	Using the Casas--Ibarra parametrization \cite{Casas:2001sr}, one can
	express the Yukawa coupling matrix $Y$ in terms of the physical
	neutrino mass matrix $m_\nu^{\rm diag}$, the PMNS matrix $U$, the
	(diagonal) RHN mass matrix $M$ and an undetermined complex orthogonal
	matrix $R$:
	\begin{equation} \label{CIY}
	Y = \frac {i} {v} U^{\star} \sqrt {m_\nu^{\rm diag}} R^T \sqrt{M}\,.
	\end{equation}
	This expression can be solved for $R$:
	\begin{align} \label{eq:R}
	(R^{T})_{ij} = -i v \frac{ (U^{T}Y)_{ij}} {\sqrt{m_i} \sqrt{M_j} }
	\Longleftrightarrow R_{ji} = -i v \frac {(U^{T}Y)_{ij}} {\sqrt{m_i}
		\sqrt{M_j} } \,.  
	\end{align}

	Since we assume $m_1 = 0$, see eq.(\ref{eq:numass2}), we actually only
	need two RHNs \cite{Ibarra:2003up}, i.e. we can set
	$M_3 \rightarrow \infty$. Since $m_2 \neq 0$ and $m_3 \neq 0$,
	eq.(\ref{eq:R}) implies that $R_{32}$ and $R_{33}$ have to vanish. On
	the other hand $R_{31}$ might not be zero since the product $m_1 M_3$
	is ill defined in our setup. We can thus write $R$ as
	\begin{equation} 
	R=\left(\begin{array}{ccc}
	R_{11} & R_{12} & R_{13} \\
	R_{21} & R_{22} &  R_{23}\\
	R_{31} & 0 & 0
	\end{array}\right)\,; \ \
	R^{T}=\left(\begin{array}{ccc}
	R_{11} & R_{21} & R_{31} \\
	R_{12} & R_{22} &  0\\
	R_{13} & R_{23} & 0
	\end{array}\right)\,,
	\end{equation}
	Since $R  R^{T} =  R^{T} R =\mathbb{1}$, one has 
	\begin{align} \label{eq:RR}
	\mathbb{1}&=
	\left(
	\begin{array}{ccc}
	R_{11}^2 + R_{12}^2 + R_{13}^2\ & R_{11} R_{21} + R_{12} R_{22}+ R_{13} R_{23}\ &
	R_{11} R_{31}  \\
	R_{11} R_{21} + R_{12} R_{22}+ R_{13} R_{23}\ & R_{21}^2 + R_{22}^2 +R_{23}^2\ &
	R_{21}R_{31} \\
	R_{11}R_{31} & R_{21} R_{31} & R_{31}^2
	\end{array}
	\right) \,; \nonumber \\
	\mathbb{1}&=
	\left(
	\begin{array}{ccc}
	R_{11}^2 + R_{21}^2 +R_{31}^2\ & R_{11} R_{12} + R_{21} R_{22}\
	& R_{11}R_{13} +  R_{21}R_{23}\\
	R_{11} R_{12} + R_{21} R_{22}\ & R_{12}^2 + R_{22}^2\
	&  R_{12}R_{13} +  R_{22}R_{23} \\
	R_{11}R_{13} +  R_{21}R_{23}\ & R_{12}R_{13} +  R_{22}R_{23}\
	& R_{13}^2+R_{23}^2
	\end{array}
	\right)\,.
	\end{align}
	The $(3,3)$ element in the first eq.(\ref{eq:RR}) implies
	$R_{31}^2=1$. The $(3,1)$ and $(3,2)$ elements of that equation then
	require $R_{11} = R_{21}=0$. The $(1,1)$ and $(2,2)$ elements in turn
	imply $R_{12}^2+R_{13}^2 = R_{22}^2+R_{23}^2 =1$, and the $(1,2)$
	element gives $R_{12} R_{22}+ R_{13} R_{23}=0$. Turning to the second
	eq.(\ref{eq:RR}), the above relations suffice to satisfy the equations
	for the $(1,1),\ (1,2)$ and $(1,3)$ elements. The $(2,2)$ and $(3,3)$
	elements imply $R_{12}^2+R_{22}^2= R_{13}^2 + R_{23}^2=1$, and the
	$(2,3)$ element requires $R_{12} R_{13}+ R_{22} R_{23}=0$. Hence we
	can write the matrix $R$ as \cite{Ibarra:2003up}
	\begin{equation} \label{Rmatrix}
	R=\left(\begin{array}{ccc}
	0 & \cos z & \sin z \\
	0 & -\sin z &  \cos z \\
	1 & 0 & 0
	\end{array}\right)\,,
	\end{equation}
	with $z$ being a complex angle.\footnote{We remind the reader that
		$\cos z = \left( {\rm e}^{iz} + {\rm e}^{-iz} \right) / 2, \
		\sin z =  \left( {\rm e}^{iz} - {\rm e}^{-iz} \right) / (2i)$ hold also
		for complex argument $z$, so that $\cos^2 z + \sin^2 z = 1$ as usual.
		Also, $(\cos z)^* = \cos (z^*)$ and $(\sin z)^* = \sin(z^*)$.}
	
	Having parameterized $R$, we can use eq.\eqref{CIY} and $m_1 = 0$ to
	derive
	\begin{equation} \label{eq:YY}
	(Y^{\dagger} Y)_{ij} = \frac{ \sqrt{M_i M_j}} {v^2} \left( m_2 R^{\star}_{i2}
	R_{j2} + m_3 R^{\star}_{i3}R_{j3}\right)\,.
	\end{equation}
	In particular, we'll need the following two quantities in the following:
	\begin{align} \label{eq:Y11}
	(Y^{\dagger} Y)_{11} = \sum_{\alpha = 1}^3 Y^\star_{\alpha 1} Y_{\alpha 1}
	& = \frac{M_1}{v^2} \left( m_2|\cos z|^2 + m_3|\sin z|^2 \right)\,;
	\\ \label{eq:Y21}
	(Y^{\dagger} Y)_{21}  & = \frac {\sqrt{M_1\, M_2}} {v^2} \left[ -m_2\sin
	z^\star \cos z + m_3 \sin z \cos z^\star \right]\,.
	\end{align}
	%
	
	\section{Baryogenesis via Leptogenesis}
	\label{leptogenesis}
	
	As noted earlier, we focus on the simplest version of leptogenesis,
	where the asymmetry is generated solely in the decay of the lightest
	RHN. The heavier RHNs cannot be produced either thermally or
	non--thermally if $M_3\gg M_2 > \Trh,\ m_\phi$. The yield of both
	thermal and non--thermal leptogenesis is then proportional to the CP
	asymmetry parameter \cite{Fong:2012buy}
	\begin{equation} \label{eq:eps}
	\epsilon_1 = \frac{1}{8 \pi} \frac {1} { \left( Y^{\dagger} Y\right)_{11}}
	\operatorname{Im} \left[ \left( Y^{\dagger} Y\right)_{2 1}^{2}\right]
	g \left( \frac {M_2^2} {M_1^2} \right)\,.
	\end{equation}
	This parameter describes the (loop--induced) difference between the
	branching ratios into CP--conjugate states, i.e. between final states
	containing a lepton and those containing an antilepton. The loop
	function $g$ in eq.(\ref{eq:eps}) is given by
	\begin{align}
	g(x) &= \sqrt{x} \left[ \frac {1} {1-x} + 1 - (1+x) \ln \left(
	\frac {1+x} {x} \right) \right]\nonumber \\
	&\simeq -\frac{3}{2}\left( \frac {1} {x}\right)^{1/2}
	- \frac {5} {6} \left( \frac {1} {x} \right)^{3/2}
	+ \mathcal{O} \left( \frac {1} {x}\right)^{5/2}
	\quad \text{for} \quad x\gg 1 \,.
	\end{align}
	From eq.(\ref{eq:Y21}) we have
	\begin{align}
	\left( Y^\dagger Y \right)_{2 1}^2 &=  \frac {M_1\, M_2} {v^4} \left[
	-m_2 \sin z^\star \cos z + m_3 \sin z \cos z^\star  \right]^2\nonumber \\
	&= \frac {M_1\, M_2} {v^4} \left[ m_2^2\,\sin^2 z^\star \,\cos^2 z
	+m^2_3 \sin^2 z \cos^2 z^\star - 2m_2 m_3 |\sin z|^2 |\cos z|^2 \right]\,,
	\end{align}
	and therefore
	\begin{align}
	\text{Im} \left[ \left( Y^\dagger Y\right)_{2 1}^2 \right]
	&= \frac {M_1\, M_2} {v^4} \text{Im} \left[ m_2^2\,\sin^2 z^\star \,\cos z^2
	+ m^2_3 \sin z^2 \cos^2 z^\star \right] \nonumber \\
	& = \frac {M_1\, M_2} {v^4} (m_3^2- m_2^2)\, \text{Im} \left( \sin^2 z \right)
	\,;
	\end{align}
	In the last step we used $\cos^2 z = 1 - \sin^2
	z$ and $\text{Im}\left( \sin^2z \sin^2 z^* \right) =
	0$. Now define $\cos^2 z = x + i\, y$, hence
	$\sin^2 z = 1- \cos^2 z = 1-x -i\, y$, with $x, y \in
	\mathbb{R}$. For $M^2_2 \gg M^2_1$ we can then derive a simple
	upper bound on $\epsilon_1$:
	\begin{align} \label{eq:epsilon1}
	\epsilon_1 &= \frac {1} {8 \pi} \frac {v^2} {M_1 \left( m_2 |\cos z|^2 
		+ m_3 |\sin z|^2 \right) } \operatorname{Im} \left[ \left( Y^\dagger Y
	\right)_{2 1}^2 \right] g \left( \frac {M_2^2} {M_1^2}\right) \nonumber \\
	& =  \frac {1} {8 \pi} \frac {v^2} {M_1} \frac {M_1\, M_2} {v^4} 
	\frac{ (m_3^2- m_2^2)\,\text{Im} \left( \sin^2 z \right)} 
	{\left( m_2|\cos z|^2 + m_3|\sin z|^2 \right)} \left( -\frac{3}{2} 
	\frac {M_1} {M_2} \right) \left[ 1 + \mathcal{O} \left( \frac{M_1^2} {M_2^2}
	\right) \right] \nonumber\\
	&\simeq -\frac {3} {16 \pi} \frac {M_1} {v^2} (m_3^2- m_2^2) 
	\frac{ \text{Im} \left( \sin^2 z \right)} {\left( m_2|\cos z|^2 + m_3|\sin z|^2
		\right)} \nonumber \\
	& = -\frac {3} {16 \pi} \frac {M_1} {v^2} (m_3^2- m_2^2)\, 
	\frac {-y} {\left( m_2 \sqrt{x^2+y^2} + m_3 \sqrt{(1-x)^2 + y^2} \right) }
	\nonumber \\
	&\lesssim \frac {3} {16 \pi} \frac {M_1} {v^2} (m_3^2- m_2^2)\, 
	\frac {y} { \left( m_2 \sqrt{1+y^2} + m_3 \sqrt{ y^2}\right)} \nonumber \\
	&\lesssim \frac {3} {16 \pi} \frac {M_1} {v^2} (m_3- m_2) 
	\simeq 10^{-5} \left( \frac {M_1} {10^{11}\ \text{GeV}} \right)
	\,.
	\end{align}
	In the fifth step we have assumed $\epsilon_1 > 0$, which is needed
	for a positive baryon asymmetry, and used $m_3 > m_2$, so that the
	denominator is minimized, and hence $\epsilon_1$ is maximized, for
	$x =1$. The penultimate step follows because the $y-$dependent
	fraction reaches its maximal value $1/(m_2+m_3)$ for
	$y \rightarrow \infty$, and in the last step we have used the
	numerical values from eq.(\ref{eq:numass2}).
	
	The Baryon Asymmetry of the Universe (BAU) today is given by
	\cite{Kolb:1990vq}
	\begin{equation}\label{etaB}
	\eta_B \equiv \frac {n_B} {n_\gamma} = \left( \frac {s} {n_\gamma} \right)_0
	\left( \frac {n_B} {\nbl} \right) \frac{\nbl}{s}
	\simeq 7 \times \frac {28} {79} Y_{B-L}\,.
	\end{equation}
	Here $n_\gamma = 2\zeta(3)T^3 / \pi^2$ and
	$s = 2\pi^2\,g_{\star\,s} T^3 / 45$ denote the photon number and
	entropy densities, respectively, $g_{\star\,s}$ being a measure of the
	number of relativistic degrees of freedom. The subscript $0$ refers to
	the current time, when $T = T_0 \simeq 2.73$~K and
	$g_{\star\,s} \simeq 3.9$ in the absence of ``dark
	radiation''. Finally, the factor $28/79$ results from the transfer of
	the primarily generated $B-L$ asymmetry to the baryon asymmetry by
	electroweak sphaleron processes.  Using the BAU value reported by
	Planck 2018, one has
	\begin{align} \label{eq:YBL_required}
	\eta_B \simeq 6 \times 10^{-10} \Rightarrow Y_{B-L} \simeq 10^{-10}\,.   
	\end{align}
	Hence one needs a $B-L$ yield $Y_{B-L} \gtrsim 10^{-10}$ in order to generate
	the observed BAU. In the subsequent subsections we will explore how a
	$B-L$ asymmetry of this magnitude can be generated in the framework of
	polynomial inflation via either thermal or non--thermal $N_1$ production.
	
	\subsection{Leptogenesis with Bosonic Reheating}
	
	In this section, we focus on the bosonic reheating scenario, where
	the inflaton dominantly decays to a pair of SM Higgs bosons. We will discuss
	thermal and non--thermal $N_1$ production in turn.
	
	\subsubsection{Thermal Leptogenesis}
	
	In the conceptually simplest case $N_1$ was dominantly generated by
	scattering reactions involving SM particles in the thermal plasma
	after reheating. The $B-L$ number yield $Y_{B-L}$ is then given by
	\cite{Buchmuller:2004nz, Fong:2012buy}:
	\begin{equation} \label{eq:YBL}
	Y_{B-L} = Y^{\text{eq}}_{N_1} \kappa_f \epsilon_1 = \frac {45} {\pi^4 \gss}
	\epsilon_1 \kappa_f \,.
	\end{equation}
	In this case $Y_{B-L}$ is mainly controlled by the scaled $N_1$
	equilibrium density, given by the numerical factor in
	eq.~\eqref{eq:YBL} (valid for $\Trh > M_1$); the CP asymmetry factor
	$\epsilon_1$; and an efficiency factor $\kappa_f$ parameterizing
	wash--out effects from inverse decays. The latter can be estimated as
	\cite{Buchmuller:2004nz}
	\begin{equation} \label{eq:kf}
	\kappa_f \simeq (2 \pm 1) \times 10^{-2} \left( \frac {0.01 \mathrm{eV}}
	{\tilde{m}_1}\right)^{1.1}\,,
	\end{equation}
	where
	$\tilde{m}_{1}= v^{2}\left( Y^\dagger Y\right)_{11} / M_1 \geq m_2
	\simeq 0.0086$~eV, so $\kappa_f \lesssim 10^{-2}$.
	
	Inserting the upper bound on $\epsilon_1$ given by \eqref{eq:epsilon1}
	and $\kappa_f \lesssim 10^{-2}$ into eq.(\ref{eq:YBL}) leads to a
	lower bound on $M_1$:
	\begin{equation}
	M_1 \gtrsim 10^{10}~\text{GeV}\,.
	\end{equation}
	Thermal leptogenesis only works if the reheating temperature
	$\Trh > M_1 \gtrsim 10^{10}~\text{GeV}$, as originally discussed in
	Ref.~\cite{Davidson:2002qv}. Specifically, (inverse) decays
	$N_1 \leftrightarrow \ell_\alpha H$ attain equilibrium at a
	temperature of a few $M_1$ (assuming the universe ever was this hot,
	of course), but will freeze out for $T$ slightly below $M_1$; note
	that the coupling involved is quite small, see \eqref{eq:Y11}. For
	bosonic inflaton decay in our model, $\Trh \geq 10^{10}$~GeV is
	possible if $\phi_0 \geq 0.4 M_P$ \cite{Drees:2021wgd}; for
	$\phi_0 \geq 10 M_P$, a reheating temperature as high as $10^{14}$~GeV
	can be achieved \cite{Drees:2022aea}. On the other hand, if the
	inflaton predominantly decays fermionically (into fermions other than
	$N_1$), $\Trh > 10^{10}$~GeV requires $\phi_0 > 6 M_P$, and the
	maximal reheat temperature consistent with radiative stability of the
	inflaton potential is about $10^{11}$~GeV \cite{Drees:2022aea}.
	Thermal leptogenesis (at least in this simple form) is therefore far
	more plausible for bosonic inflaton decays in our model.
	
	\subsubsection{Non--thermal Leptogenesis via Inflaton Three--Body Decay}
	\label{sec:3body}
	
	Having shown that standard thermal leptogenesis is possible in our
	scenario, here we point out that, at least in the large field version
	of the model, non--thermal leptogenesis is also possible, even if
	the inflaton primarily decays bosonically, specifically into pairs of
	SM Higgs bosons. As far as we are aware this is a novel variant of
	leptogenesis.
	
	The main observation is that for dominant $\phi \rightarrow HH$
	decays, the inflaton can also undergo three--body decay
	$\phi \to H \ell N_1$ as long as $M_1 < m_\phi$. This can happen if
	one of the Higgs bosons in the final state is (far) off-shell, and
	transitions into $N_1$ plus an SM lepton $\ell$; see
	fig.~\ref{threedecay} in the Appendix. This scenario is particularly
	interesting for relatively small inflaton couplings, i.e. for
	$\Trh < M_1$, where the usual thermal leptogenesis is not possible.
	We emphasize that the coupling that allows this three--body decay is
	automatically present in the type I see--saw model.  Specifically, the
	branching ratio for the three--body decay is proportional to
	$\sum_\alpha Y^*_{\alpha 1} Y_{\alpha 1}$, which we evaluated in
	eq.(\ref{eq:Y11}). Note that this precisely cancels the denominator of
	$\epsilon_1$, see eq.\eqref{eq:epsilon1}.  As shown in the Appendix,
	the resulting $B-L$ yield is given by:
	\begin{align}\label{eq:inflaton_three_body_decay}
	Y_{B-L}
	&=\left[ \frac {3} {4} \frac {\Trh} {m_\phi}\, \text{BR} (\phi \to \ell N_1
	H )\right] \, \epsilon_1\nonumber \\
	& \simeq  -4.8 \times 10^{-11} \left( \frac {m_\phi} {10^{11}\ \text{GeV}}
	\right) \left( \frac {\Trh} {10^{11}\ \text{GeV}} \right) {\rm Im}
	\left(\sin^2 z\right) \cdot \sqrt{\gamma} f(\gamma)\,.
	\end{align} 
	The function $f(\gamma)$ appears in the calculation of the branching
	ratio for the three--body decay; it is given by
	\begin{equation} \label{eq:f}
	f(\gamma) = \sqrt{\gamma} \left[ \gamma^2 +4 \gamma - 5 -2(1+2\gamma)\ln
	\gamma \right]\,,
	\end{equation}
	with $\gamma \equiv M_1^2/m_\phi^2$. The expression in square
	parenthesis results from integrating the squared matrix element over
	the three--body phase space, whereas the factor $\sqrt{\gamma}$
	results from the squared $N_1$ Yukawa coupling. Note that
	$0<\gamma < 1$; for $\gamma \ll 1$, the phase space factor is
	dominated by the term $-2 \ln \gamma$. Of course,
	$f(\gamma) \rightarrow 0$ for $\gamma \to 1$ or equivalently
	$M_1 \to m_\phi$, since this corresponds to the closing of phase
	space; specifically, $f(\gamma) \simeq (1-\gamma)^4/6$ as
	$\gamma \rightarrow 1$. We note that $\sqrt{\gamma} f(\gamma)$ reaches
	its maximal value of $0.095$ at $\gamma = 0.081$. Finally, in
	eq.\eqref{eq:inflaton_three_body_decay} we have assumed that no
	wash--out occurs. This is an oversimplification if $\Trh$ is only
	slightly below $M_1$.
	
	\begin{figure}[ht]
		\def\sepf{0.55}
		\centering
		\includegraphics[scale=\sepf]{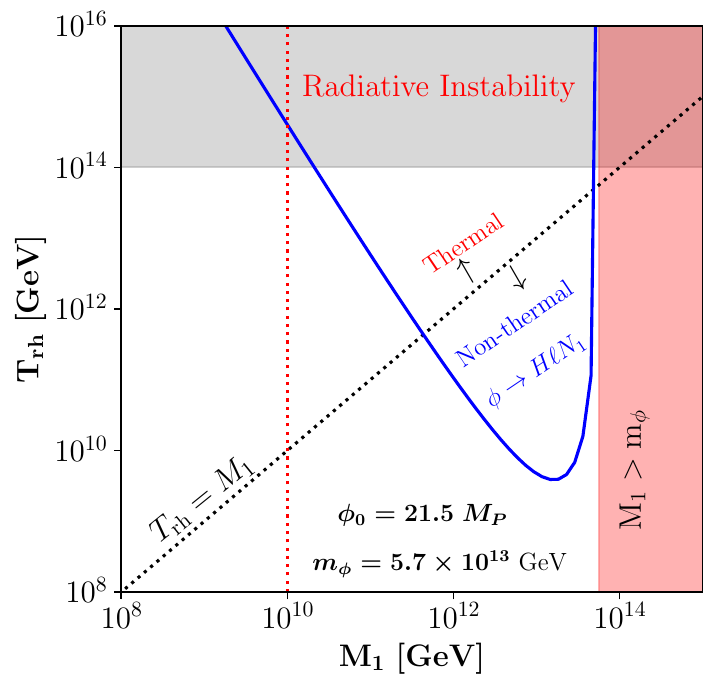}
		\includegraphics[scale=\sepf]{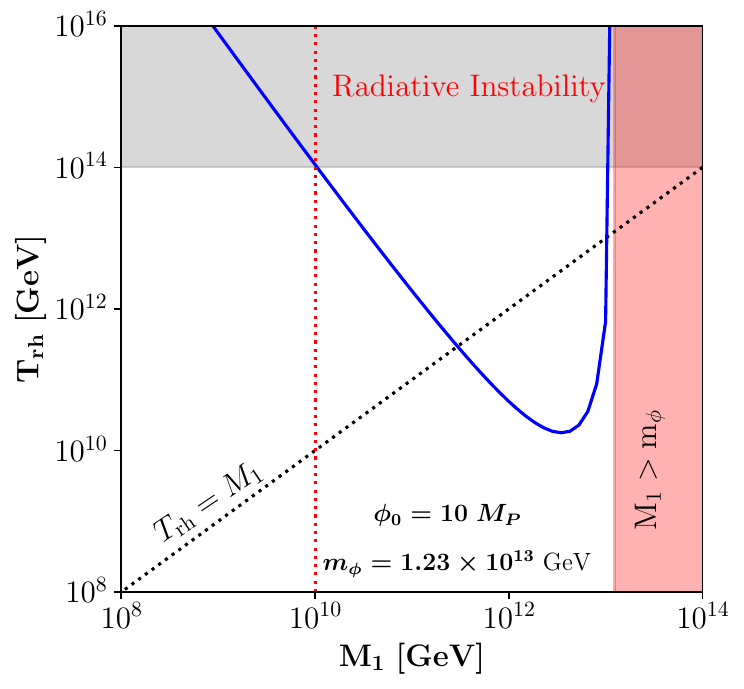}\\
		\includegraphics[scale=\sepf]{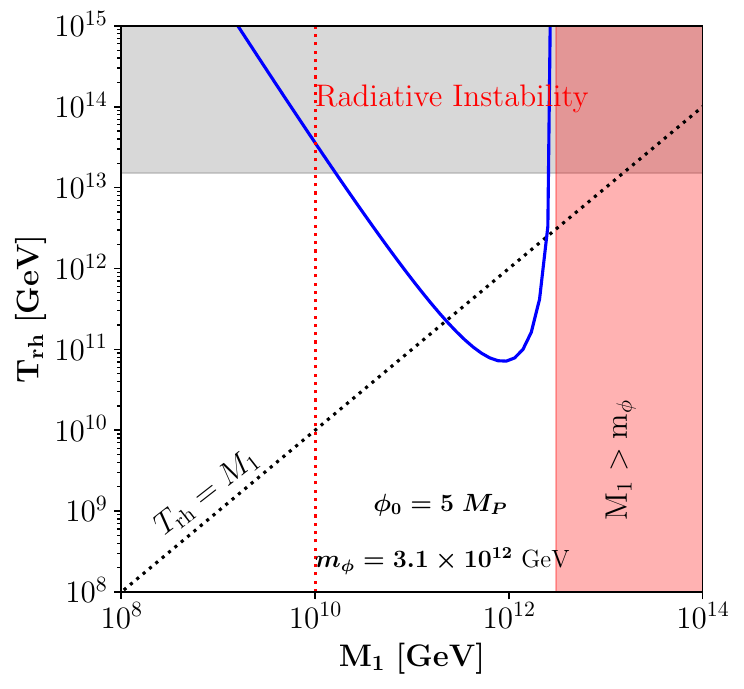}
		\includegraphics[scale=\sepf]{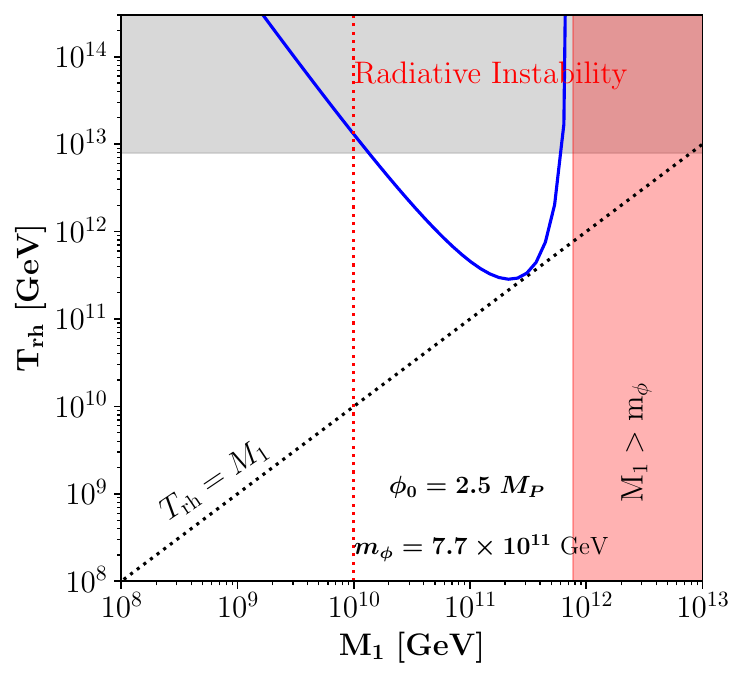}
		\caption {$\Trh$ as function of $M_1$ that yields the observed Baryon
			asymmetry via inflaton three body decays (blue line) from
			eq.\eqref{eq:inflaton_three_body_decay} with
			${\rm Im}\left(\sin^2 z\right) = 1$, for four different values of
			$\phi_0$. In the gray shaded region at the top loop corrections
			dominate the slope of the potential at $\phi_0$, invalidating our
			analysis of inflation. In the pink shaded region at the right
			$\phi \rightarrow H  \ell N_1$ decays are kinematically
			forbidden. The dotted vertical red line indicates the minimal value
			of $M_1$ allowing thermal leptogenesis; the latter is only possible
			above the dotted black lines.}
		\label{fig:phiHLN1}
	\end{figure} 
	
	We saw in eq.\eqref{eq:YBL_required} that $\YBL \gtrsim 10^{-10}$ is
	required in order to reproduce the observed BAU. The blue curves in
	Figs.~\ref{fig:phiHLN1} show the minimal value of $\Trh$ that allows
	to satisfy this condition, using
	eq.\eqref{eq:inflaton_three_body_decay}; for definiteness we have
	taken ${\rm Im} \left(\sin^2 z\right) = 1$. The upper left panel is for
	$\phi_0=21.5 \ M_P$, leading to
	$m_\phi \simeq 5.7 \times 10^{13}$~GeV; this large value of $\phi_0$,
	and correspondingly heavy inflaton, saturates the current upper bound
	on $r$. We exclude the gray shaded region, where
	$\Trh \gtrsim 2 \times 10^{14}$~GeV, since here the Higgs loop will
	spoil the flatness of inflaton potential so that the inflationary
	predictions, which are based on the tree--level potential, cannot be
	trusted. The black dotted line corresponds to $\Trh =M_1$, above which
	$N_1$ particles thermalize, leading to thermal leptogenesis as
	investigated above. Note that when $\Trh >M_1$ the inflaton three-body
	decay could still generate $N_1$ particles, which however would not
	change their number density if they thermalize. The red dotted line
	corresponds to $M_1\simeq 10^{10}~\text{GeV}$, denoting the lower
	bound on $M_1$ for thermal leptogenesis.
	
	Below the black dotted line, where $M_1 > \Trh$, thermal leptogenesis
	is suppressed. In such case we find that the non--thermal channel via
	inflaton three--body decays might still be able to source the required
	BAU, as depicted by the blue line. When $M_1 \to m_\phi$, the
	three-body decay is phase space closed, explaining why the blue curves
	suddenly go up.
	
	We also show examples with smaller $\phi_0$ in Fig.~\ref{fig:phiHLN1}:
	upper right panel $\phi_0 =10\, M_P$, lower left panel
	$\phi_0 =5\, M_P$ and lower right panel $\phi_0 =2.5\, M_P$.  For
	$\phi_0 \lesssim 2.5\, M_P$, ${\rm Im}(\sin^2 z) > 1$ would be
	required, which would require some cancellation to occur in
	eq.\eqref{mnu1}.
	
	Several comments are in order before we close this section. First,
	non--thermal leptogenesis via inflaton three--body decay investigated
	in this section is also viable for other inflation models (e.g. the
	Starobinsky \cite{Starobinsky:1980te} and $\alpha$-attractor models
	\cite{Kallosh:2013hoa}) as long as the inflaton mass satisfies
	$m_\phi \gtrsim 10^{12}~\text{GeV}$ and $\Trh \gtrsim
	10^{10}$~GeV. Second, it might also work in scenarios with a later
	epoch of matter domination by some very massive particle, e.g. a
	modulus field \cite{Drees:2017iod} which predominantly decays into a
	pair of SM Higgs bosons. The above bound on $m_\phi$ should then be
	satisfied by this modulus field, and $\Trh$ refers to the temperature
	at the end of the modulus matter dominated epoch. Finally, for
	$m_\phi > 2 M_1$ the inflaton can also decay into pairs of RHNs,
	either via four--body decays, or via loop-induced two-body decays as
	commented in the end of Sec.~\ref{setup}. However, these processes
	have (even) smaller branching ratios than the three--body final state,
	which will therefore dominate non--thermal $N_1$ production in our
	bosonic reheating scenario.
	
	\subsection{Non--thermal Leptogenesis with Fermionic Reheating}
	\label{sec:non_thermal_lep}
	
	\begin{figure}[ht]
		\def\sepf{0.55}
		\centering
		\includegraphics[scale=\sepf]{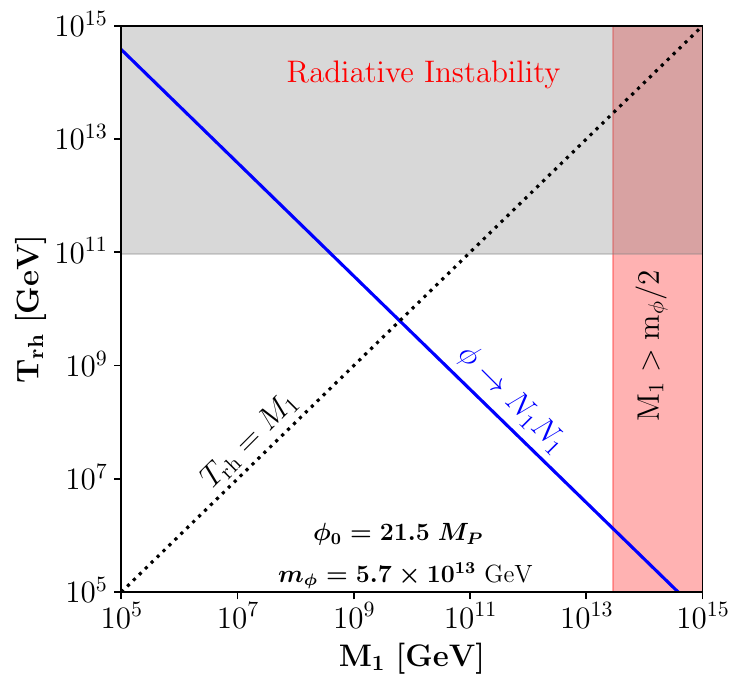}
		\includegraphics[scale=\sepf]{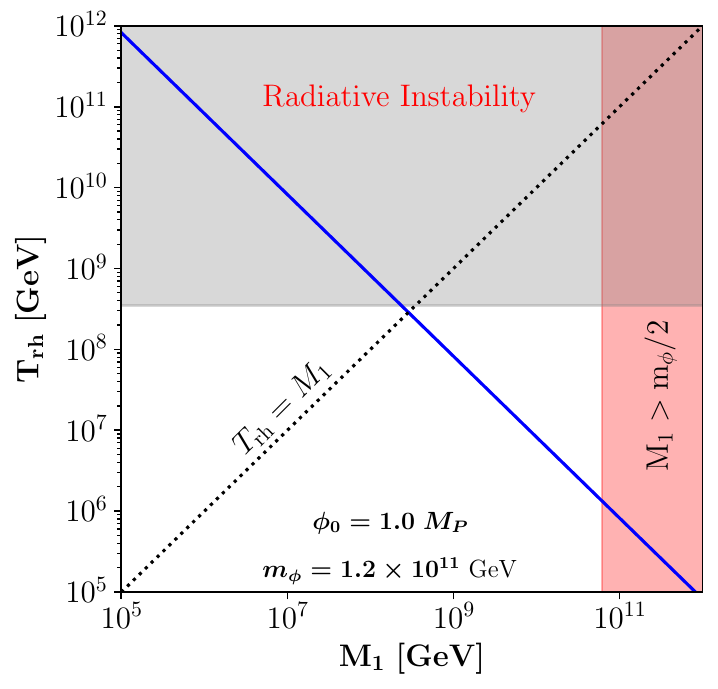}
		\includegraphics[scale=\sepf]{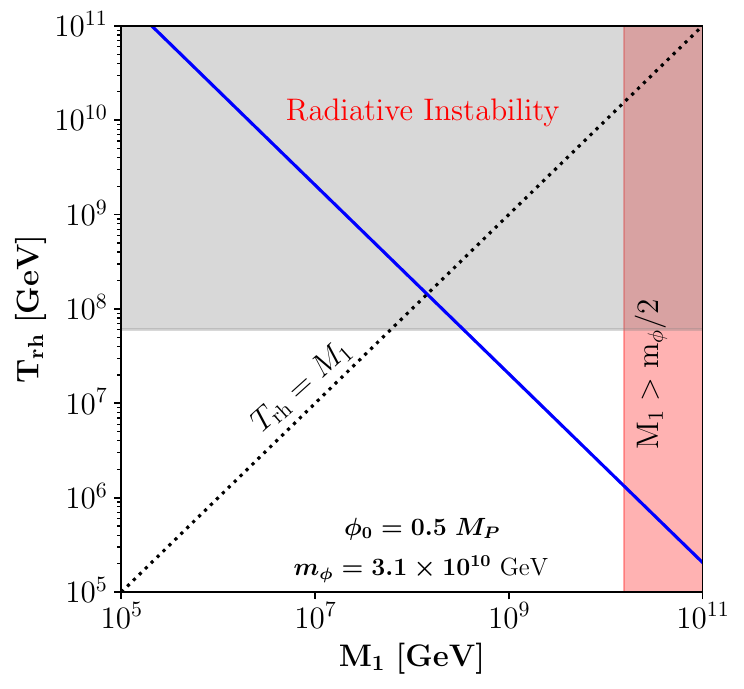}
		\includegraphics[scale=\sepf]{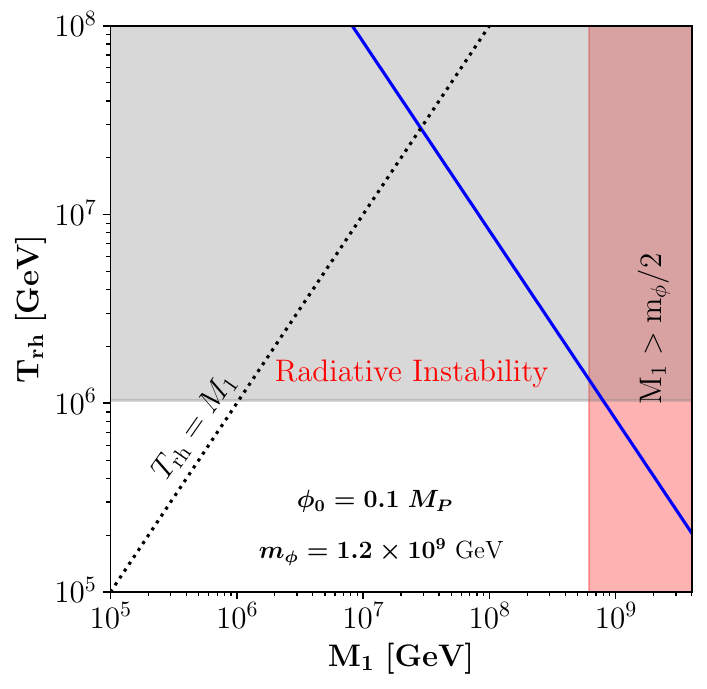}
		\caption {Parameter space for non--thermal leptogenesis if the
			inflaton predominantly decays into two $N_1$, for four different
			values of $\phi_0$ and hence of $m_\phi$. Above the blue lines the
			correct BAU can be generated non--thermally, see eq.\eqref{YBL2}. In
			the gray shaded regions our tree--level analysis of inflation is not
			reliable, while in the pink shaded regions
			$\phi \rightarrow N_1 N_1$ decays are kinematically
			forbidden. Finally, below the dotted black lines $N_1$ does not
			attain thermal equilibrium.}
		\label{fig:non--thermal}
	\end{figure}
	
	In this section we assume that RHNs are produced via inflaton two body
	decays, $\phi \to N_1 N_1$ \cite{Lazarides:1990huy, Giudice:1999fb,
		Asaka:1999jb, Senoguz:2003hc, Hahn-Woernle:2008tsk, Antusch:2010mv};
	these decays also reheat the universe. We saw above that in our
	inflationary scenario $\Trh$ is too low for thermal leptogenesis if
	the inflaton decays fermionically, except for a very limited region of
	parameter space at large $\phi_0$. Here we show that, even in our very
	simple type I see--saw model, the parameter space for leptogenesis is
	greatly expanded by this non--thermal $N_1$ production channel.
	
	We will again assume that washout effects can be neglected if
	$\Trh < M_1$\footnote{The bulk  of the production for $N_1$ occurs  when $T=T_{\text{rh}}$, at which the washout effect is Boltzmann suppressed if $\Trh < M_1$.}. The $B-L$ number yield after reheating can then simply
	be estimated as \cite{ Kolb:1990vq, Asaka:1999jb}:
	\begin{equation} \label{YBL2}
	Y_{B-L} = \frac {\nbl} {s} = \left[ \frac {3} {2} \frac {\Trh} {m_\phi}\,
	\text{BR} (\phi \to N_1N_1) \right] \, \epsilon_1\,.
	\end{equation}
	The expression between the square parentheses denotes the yield of
	$N_1$ from inflaton decay, allowing for a general branching ratio
	$\text{BR}$ for $\phi \rightarrow N_1 N_1$; in our scenario of
	fermionic reheating with a hierarchical RHN spectrum, this branching
	ratio is unity, $\text{BR}= 1$.
	
	The allowed parameter space $(\Trh, M_1)$ is shown in
	Fig.~\ref{fig:non--thermal}. Here we consider $\phi_0 =21.5\, M_P$
	(upper left), corresponding to the largest allowed value compatible
	with the latest CMB experiments; $\phi_0 =1.0\,M_P$ (upper right),
	$\phi_0 =0.5\,M_P$ (lower left), and $\phi_0 =0.1\,M_P$ (lower
	right). The gray regions are again excluded since radiative
	corrections from the $N_1$ loop would spoil the flatness of the
	inflaton potential near $\phi_0$. In the pink regions, $M_1> m_\phi/2$
	hence $\phi \rightarrow N_1 N_1$ is kinematically
	impossible.\footnote{A small sliver of this region, with $M_1$ only
		slightly above $m_\phi/2$, may still be acceptable. The dominant
		inflaton decay mode in this region is $\phi \rightarrow N_1 H \ell$
		via off--shell $N_1$ exchange. However, the inflaton decay width,
		and hence $\Trh$, drops very quickly in this region; eq.\eqref{YBL2}
		shows that this also suppresses the generated BAU.} The blue line
	gives $Y_{B-L} = 10^{-10}$ if the upper bound \eqref{eq:epsilon1} on
	$\epsilon_1$ is used in eq.\eqref{YBL2}, i.e. below these lines the
	generated net lepton number is always too small. Finally, the black
	dotted lines again show $\Trh = M_1$, i.e. below these lines wash--out
	effects are small but above these lines $N_1$ thermalizes.
	
	As remarked earlier, in a small region of parameter space with large
	$\phi_0$ and $\Trh > M_1$ (above the dotted black lines) thermal
	leptogenesis can occur. We note that the blue solid and black dotted
	lines intersect at
	\begin{equation}
	M_1^{\prime} = 10^3\ {\text{GeV}}\;\cdot \sqrt{ \frac {2}{3} \frac {m_\phi}
		{\text{GeV}} } \,.
	\end{equation}
	Non--thermal leptogenesis can only occur for $M_1 > M_1^{\prime}$. We see that
	for $\phi_0 < M_P$ this is automatically satisfied by all points on
	the blue line where radiative corrections to the inflaton potential
	are under control. In this ``small field'' inflationary scenario, the
	maximal reheating temperature
	$T_{\rm rh}^{\rm max} \propto \phi_0^{5/2}$ for fermionic reheating
	\cite{Drees:2021wgd}. Moreover, eqs. \eqref{YBL2} and
	\eqref{eq:epsilon1} show that $Y_{B-L}$ is maximal for maximal
	$M_1 \simeq m_\phi/2$, hence the maximal
	$Y_{B-L} \propto T_{\rm rh}^{\rm max} \propto \phi_0^{5/2}$. In
	particular, for $\phi_0 \leq 0.1 M_P$, even non--thermal leptogenesis
	becomes impossible in our simple neutrino mass model.
	
	Altogether the allowed range of $\phi_0$ leading to successful
	leptogenesis via $\phi \rightarrow N_1 N_1$ decays, compatible with
	both CMB observations and neutrino oscillation data, is
	\begin{align}
	0.1\, M_P\lesssim \phi_0 \lesssim 21.5\, M_P\,.
	\end{align}
	In addition, the reheating temperature has to lie in the range
	\begin{equation} \label{ps1}
	10^{6}~\text{GeV} \lesssim \Trh \lesssim 10^{10}~\text{GeV}\,,
	\end{equation} 
	while the mass of the lightest RHN should satisfy
	\begin{equation} \label{eq:Mrange}
	10^{8}~\text{GeV} < M_1 \lesssim 3 \cdot 10^{13}~\text{GeV}\,.
	\end{equation}
	%
	
	\section{Summary and Conclusions}
	\label{sum}
	
	Polynomial inflation is a very simple and predictive model. It fits
	current cosmic microwave background (CMB) data well, and makes
	testable predictions for the running of the spectral index and the
	tensor to scalar ratio.
	
	A further necessary condition for a successful cosmological model is
	that it offers an explanation for the observed baryon asymmetry of
	the Universe (BAU). In this work we therefore revisit leptogenesis in
	the framework of polynomial inflation. We assume a type--I see--saw
	model with massless lightest neutrino, and therefore only two very
	heavy right--handed neutrinos. We show that even in this simple
	ansatz, in some part of the parameter space both the observed BAU and
	the neutrino oscillation data can be explained.
	
	To that end we consider two different perturbative reheating
	scenarios. If the inflaton decays bosonically, the reheating
	temperature in our model can be as high as
	$\Trh \sim 10^{14}~\text{GeV}$, compatible with ``plain vanilla''
	$N_1$ thermal leptogenesis. We also point out, for the first time,
	that even for $\Trh < M_1$, where thermal leptogenesis does not work,
	inflaton three--body decays of the inflaton into an SM Higgs boson, a
	RHN $N_1$ and an SM doublet lepton can lead to successful
	leptogenesis. As long as it is kinematically allowed, this decay will
	always occur if the primary inflaton decay mode is into two SM Higgs
	bosons; the required couplings are automatically present in the
	see--saw neutrino mass model, and can be expressed in terms of
	neutrino masses and a single complex mixing angle. We find that this
	three--body decay scenario is viable for inflaton mass
	$m_\phi \gtrsim 10^{12}~\text{GeV}$. In the polynomial inflation
	scenario this corresponds to $\phi_0 \gtrsim 2.5~M_P$, but this novel
	mechanism should also be applicable to other models of large--scale
	inflation.
	
	We also investigate the case where the inflaton primarily decays into
	two $N_1$. This fermionic decay mode leads to somewhat lower maximal
	reheating temperature. Thermal leptogenesis therefore works only in a
	rather small region of parameter space, but this inflaton decay mode
	allows non--thermal leptogenesis, if $\phi_0 \gtrsim 0.1~M_P$,
	$M_1 \gtrsim 3 \cdot 10^8$~GeV and reheating temperature
	$\Trh \gtrsim 10^{6}~\text{GeV}$.
	
	In conclusion, we have shown that a model with a renormalizable
	inflaton potential, in combination with a simple see--saw model with
	hierarchical masses of the right--handed neutrinos, allows successful
	thermal or non--thermal leptogenesis if inflation occurred at a field
	value $\phi \simeq \phi_0 \gtrsim 0.1 M_P$, corresponding to an Hubble
	parameter during inflation $H_I \gtrsim 2 \cdot 10^7$~GeV. Deviating
	from our assumption of a very hierarchical spectrum of right--handed
	neutrinos should allow to construct viable models operating at much
	smaller energy scales. For example, it is known that thermal
	``resonant'' leptogenesis, where two RHNs are nearly degenerate, can
	work at the TeV scale \cite{Pilaftsis:2003gt}. It would be interesting
	to investigate this quantitatively.
	
	\section*{Acknowledgments}
	
	Y.X. acknowledges the support from the Cluster of Excellence
	``Precision Physics, Fundamental Interactions, and Structure of
	Matter'' (PRISMA$^+$ EXC 2118/1) funded by the Deutsche
	Forschungsgemeinschaft (DFG, German Research Foundation) within the
	German Excellence Strategy (Project No. 390831469).
	
	\appendix 
	
	\section{Inflaton Three--Body Decay into Right Handed Neutrino}
	\label{inflaton_three_body_decay}
	
	In this appendix, we first calculate the branching ratio for inflaton
	three--body decays $\phi \to H\, \ell\, N_1$. Then we estimate the
	$B-L$ yield generated from the subsequent decay of $N_1$. 

	\begin{figure}[ht]
		\def\sepf{0.8}
		\centering
		\includegraphics[scale=\sepf]{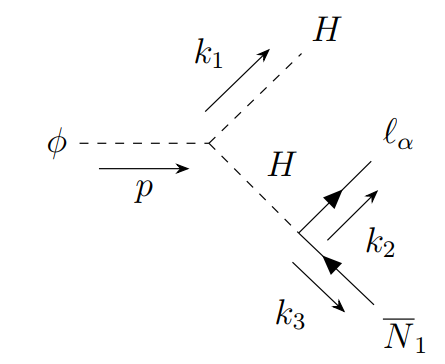}
		\caption{$N_1$ production via inflaton three--body decay.}
		\label{threedecay}
	\end{figure} 

	Our basic assumption is that the main inflaton decay mode is into a
	pair of SM Higgs bosons via the coupling $\lambda_{12}$. The
	three--body decay shown in fig.~\ref{threedecay} is then automatically
	allowed, as long as $M_1 < m_\phi$. The RHN $N_1$ couples to the SM
	Higgs and lepton doublet through the Yukawa coupling
	$Y_{\alpha 1}\, \bar{\ell}_\alpha \tilde{H} N_1 $, which is required
	for the explanation of the light neutrino masses.
	
	We label the momenta for particles involved as follows:
	\begin{align}
	\phi(p) \to H(k_1)\,H(q)\to H(k_1) \,\ell(k_2) N_1(k_3)\,, \nonumber
	\end{align}
	The matrix element for a fixed lepton flavor $\alpha$ is given by
	\begin{equation}
	i \mathcal{M}_{\alpha} = 2i\, \lambda_{12}\,\frac{i}{q^2 - m_H^2}\, Y_{\alpha\,1}
	\bar{u}(k_2) \text{v}(k_3)\,.
	\end{equation}
	After squaring and summing over lepton flavor and the $SU(2)$ index,
	this yields:
	\begin{align} \label{eq:Msquare}
	|\mathcal{M}|^2 &= \sum_\alpha 4\,\lambda^2_{12}\left( Y^\star_{\alpha\,1} Y_{\alpha\,1}
	\right) \left( 4k_2 \cdot k_3 - 4 M_1 m_{\ell_\alpha}\right)
	\frac {1} {q^4} \nonumber\\
	& \simeq 16\, \lambda^2_{12} k_2 \cdot k_3 \frac {1} {q^4} \sum_{\alpha}
	\left( Y^{\star}_{\alpha\,1}Y_{\alpha\,1} \right)\,.
	\end{align}
	In the second step we have neglected the small lepton mass. Note also
	that we assume unbroken $SU(2)$, hence all four (real) components of the Higgs
	doublet, and both charged and (light) neutral leptons, contribute. 
	
	The three--body phase space integral can be written as (see e.g. Appendix B of
	\cite{Barger:1987nn}):
	\begin{equation}\label{eq:ps_integral}
	\int d\Pi_3  =  \frac{m_\phi^2}{128\pi^3} \int^{1-\gamma}_0 d x_1
	\int^{1-\frac{\gamma}{1-x_1}}_{1-x_1-\gamma} d x_2\,,
	\end{equation}
	where $x_i = 2 E_i / m_\phi\,, \ i = 1,2$, $E_1$ being the energy of
	the Higgs boson and $E_2$ that of the charged lepton in the inflaton
	rest frame, and $\gamma = M_1^2/m_\phi^2$. We have again neglected all
	masses in the final state except for that of the heavy RHN $N_1$.  The
	product of $4-$vectors appearing in the squared matrix element
	\eqref{eq:Msquare} is related to $x_1$ via
	$2 k_2 \cdot k_3 = m_\phi^2(1-x_1) -M_1^2$.
	
	Inserting eq.~\eqref{eq:Msquare} into eq.\eqref{eq:ps_integral} allows
	to compute the three--body decay rate:
	\begin{align} 
	\Gamma_{\text{3-body}}\equiv \Gamma_{\phi \to \ell H N_1}
	&=  \frac{1}{2m_\phi} \int d\Pi_3 |\mathcal{M}|^2 \nonumber \\
	& =  \frac {\lambda^2_{12}} {32\pi^3\, m_\phi} \int^{1-\gamma}_0 d x_1
	\int^{1-\frac{\gamma}{1-x_1}}_{1-x_1-\gamma} d x_2 \frac { \, (1-x_1) -\gamma}
	{(1-x_1)^2} \,\sum_{\alpha} \left( Y^\star_{\alpha\,1}Y_{\alpha\,1} \right)
	\nonumber \\
	& = \frac {\lambda^2_{12}} {32\pi^3\, m_\phi} \sum_\alpha \left(
	Y^\star_{\alpha\,1} Y_{\alpha\,1} \right) \frac {1} {2} \left[
	\gamma^2 + 4\gamma - 5 -2(1+2\gamma) \ln\gamma\right]\,.
	\end{align}
	The branching ratio for the channel $\phi \to \ell N_1 H$ is therefore
	\begin{align}
	\text{BR}(\phi \to \ell N_1 H )
	&= \frac {\Gamma_{\text{3-body}}} {\Gamma_{\text{3-body}}
		+ \Gamma_{\phi \to H^{\dagger} H}} \simeq \frac{\Gamma_{\text{3-body}}}
	{ \Gamma_{\phi \to H^{\dagger} H}}  \nonumber \\
	& = \frac {m_\phi} {8\pi^2 v^2} \left( m_2|\cos z|^2 + m_3|\sin z|^2 \right)
	\cdot f(\gamma)\,,
	\end{align}
	where we have used
	$\Gamma_{\phi \to H^{\dagger} H}= \lambda_{12}^2 / (8\pi\, m_\phi)$,
	and eq.~\eqref{eq:Y11} for
	$\sum_{\alpha} \left( Y^{\dagger}_{\alpha\,1}Y_{\alpha\,1}
	\right)$. The function $f(\gamma)$ is defined as:
	\begin{align}
	f(\gamma) = \sqrt{\gamma}  \left[\gamma^2 +4\gamma - 5 - 2(1+2\gamma)
	\ln \gamma \right]\,.
	\end{align}
	It peaks at $\gamma \simeq 0.0177$ with $f(\gamma)^{\text{max}} \simeq 0.456$. 
	
	If washout effects can be neglected, i.e. for $M_1 > \Trh$, the
	generated $B-L$ yield is given by:\footnote{Note the suppression by a
		factor of $1/2$ relative to eq.~\eqref{YBL2}, which is due to the fact
		that now only a single $N_1$ is produced in each three--body inflaton
		decay.}
	\begin{align}
	Y_{B-L}
	&= \left[ \frac {3} {4} \frac {\Trh} {m_\phi}\, \text{BR} (\phi \to \ell N_1 H)
	\right] \, \epsilon_1\nonumber \\
	& = \frac {3} {4} \frac {\Trh} {m_\phi} \left[ \frac {m_\phi} {8\pi^2 v^2}
	\left( m_2|\cos z|^2 + m_3|\sin z|^2 \right) \cdot f(\gamma)\right]
	\nonumber \\
	& \hspace*{5mm} \cdot
	\left[ -\frac {3} {16 \pi} \frac {M_1} {v^2} (m_3^2- m_2^2)
	\frac {\text{Im} \left( \sin^2 z \right) } {\left( m_2|\cos z|^2
		+ m_3|\sin z|^2 \right) } \right] \nonumber \\
	&= -\frac {9} {512 \pi^2} \frac {M_1\, \Trh} {v^4} (m_3^2- m_2^2)
	\text{Im} \left( \sin^2 z \right)  \cdot f(\gamma) \nonumber \\
	& \simeq -4.8 \times 10^{-11} \left( \frac {m_\phi} {10^{11}\ \text{GeV}} \right)
	\left( \frac {\Trh} {10^{11}\ \text{GeV}} \right) \text{Im} \left(
	\sin^2 z \right)  \cdot \sqrt{\gamma} f(\gamma)\,.
	\end{align} 
	In the third line we have utilized eq.~\eqref{eq:epsilon1}.
	
	Before closing this section, we note that RHNs could also be produced
	in the scattering of an inflaton and a daughter particle,
	$\phi H \to \ell N_1$ and $\phi\, \ell \to H N_1$. For $m_\phi > M_1$
	this is possible even if the daughter particle in the initial state is
	soft, e.g. after thermalization with low reheating temperature. The
	rate for these scattering reactions involves the same product of
	couplings that appears in the three--body decay width. However the
	rate for inflaton scattering on a thermalized daughter particle is
	suppressed by a factor $T/m_\phi$ compared to the three--body decay
	rate; the rate for inflaton scattering on daughter particles before
	they thermalize is even smaller. We have therefore neglected these
	$2\to 2$ scattering reactions.
	
	\bibliographystyle{JHEP}
	\bibliography{biblio}

\providecommand{\href}[2]{#2}\begingroup\raggedright\begin{thebibliography}{10}

\bibitem{Planck:2018jri}
{\scshape Planck} collaboration, \emph{{Planck 2018 results. X. Constraints on
  inflation}}, \href{https://doi.org/10.1051/0004-6361/201833887}{\emph{Astron.
  Astrophys.} {\bfseries 641} (2020) A10}
  [\href{https://arxiv.org/abs/1807.06211}{{\ttfamily 1807.06211}}].

\bibitem{BICEP:2021xfz}
{\scshape BICEP, Keck} collaboration, \emph{{Improved Constraints on Primordial
  Gravitational Waves using Planck, WMAP, and BICEP/Keck Observations through
  the 2018 Observing Season}},
  \href{https://doi.org/10.1103/PhysRevLett.127.151301}{\emph{Phys. Rev. Lett.}
  {\bfseries 127} (2021) 151301}
  [\href{https://arxiv.org/abs/2110.00483}{{\ttfamily 2110.00483}}].

\bibitem{Drees:2021wgd}
M.~Drees and Y.~Xu, \emph{{Small field polynomial inflation: reheating,
  radiative stability and lower bound}},
  \href{https://doi.org/10.1088/1475-7516/2021/09/012}{\emph{JCAP} {\bfseries
  09} (2021) 012} [\href{https://arxiv.org/abs/2104.03977}{{\ttfamily
  2104.03977}}].

\bibitem{Drees:2022aea}
M.~Drees and Y.~Xu, \emph{{Large field polynomial inflation: parameter space,
  predictions and (double) eternal nature}},
  \href{https://doi.org/10.1088/1475-7516/2022/12/005}{\emph{JCAP} {\bfseries
  12} (2022) 005} [\href{https://arxiv.org/abs/2209.07545}{{\ttfamily
  2209.07545}}].

\bibitem{Munoz:2016owz}
J.B.~Mu\~noz, E.D.~Kovetz, A.~Raccanelli, M.~Kamionkowski and J.~Silk,
  \emph{{Towards a measurement of the spectral runnings}},
  \href{https://doi.org/10.1088/1475-7516/2017/05/032}{\emph{JCAP} {\bfseries
  05} (2017) 032} [\href{https://arxiv.org/abs/1611.05883}{{\ttfamily
  1611.05883}}].

\bibitem{Bernal:2021qrl}
N.~Bernal and Y.~Xu, \emph{{Polynomial inflation and dark matter}},
  \href{https://doi.org/10.1140/epjc/s10052-021-09694-5}{\emph{Eur. Phys. J. C}
  {\bfseries 81} (2021) 877}
  [\href{https://arxiv.org/abs/2106.03950}{{\ttfamily 2106.03950}}].

\bibitem{Fukugita:1986hr}
M.~Fukugita and T.~Yanagida, \emph{{Baryogenesis Without Grand Unification}},
  \href{https://doi.org/10.1016/0370-2693(86)91126-3}{\emph{Phys. Lett. B}
  {\bfseries 174} (1986) 45}.

\bibitem{Buchmuller:2004nz}
W.~Buchmuller, P.~Di~Bari and M.~Plumacher, \emph{{Leptogenesis for
  pedestrians}}, \href{https://doi.org/10.1016/j.aop.2004.02.003}{\emph{Annals
  Phys.} {\bfseries 315} (2005) 305}
  [\href{https://arxiv.org/abs/hep-ph/0401240}{{\ttfamily hep-ph/0401240}}].

\bibitem{Covi:1996wh}
L.~Covi, E.~Roulet and F.~Vissani, \emph{{CP violating decays in leptogenesis
  scenarios}}, \href{https://doi.org/10.1016/0370-2693(96)00817-9}{\emph{Phys.
  Lett. B} {\bfseries 384} (1996) 169}
  [\href{https://arxiv.org/abs/hep-ph/9605319}{{\ttfamily hep-ph/9605319}}].

\bibitem{Fong:2012buy}
C.S.~Fong, E.~Nardi and A.~Riotto, \emph{{Leptogenesis in the Universe}},
  \href{https://doi.org/10.1155/2012/158303}{\emph{Adv. High Energy Phys.}
  {\bfseries 2012} (2012) 158303}
  [\href{https://arxiv.org/abs/1301.3062}{{\ttfamily 1301.3062}}].

\bibitem{Lazarides:1990huy}
G.~Lazarides and Q.~Shafi, \emph{{Origin of matter in the inflationary
  cosmology}}, \href{https://doi.org/10.1016/0370-2693(91)91090-I}{\emph{Phys.
  Lett. B} {\bfseries 258} (1991) 305}.

\bibitem{Giudice:1999fb}
G.F.~Giudice, M.~Peloso, A.~Riotto and I.~Tkachev, \emph{{Production of massive
  fermions at preheating and leptogenesis}},
  \href{https://doi.org/10.1088/1126-6708/1999/08/014}{\emph{JHEP} {\bfseries
  08} (1999) 014} [\href{https://arxiv.org/abs/hep-ph/9905242}{{\ttfamily
  hep-ph/9905242}}].

\bibitem{Asaka:1999yd}
T.~Asaka, K.~Hamaguchi, M.~Kawasaki and T.~Yanagida, \emph{{Leptogenesis in
  inflaton decay}},
  \href{https://doi.org/10.1016/S0370-2693(99)01020-5}{\emph{Phys. Lett. B}
  {\bfseries 464} (1999) 12}
  [\href{https://arxiv.org/abs/hep-ph/9906366}{{\ttfamily hep-ph/9906366}}].

\bibitem{Asaka:1999jb}
T.~Asaka, K.~Hamaguchi, M.~Kawasaki and T.~Yanagida, \emph{{Leptogenesis in
  inflationary universe}},
  \href{https://doi.org/10.1103/PhysRevD.61.083512}{\emph{Phys. Rev. D}
  {\bfseries 61} (2000) 083512}
  [\href{https://arxiv.org/abs/hep-ph/9907559}{{\ttfamily hep-ph/9907559}}].

\bibitem{Senoguz:2003hc}
V.N.~Senoguz and Q.~Shafi, \emph{{GUT scale inflation, nonthermal leptogenesis,
  and atmospheric neutrino oscillations}},
  \href{https://doi.org/10.1016/j.physletb.2003.12.020}{\emph{Phys. Lett. B}
  {\bfseries 582} (2004) 6}
  [\href{https://arxiv.org/abs/hep-ph/0309134}{{\ttfamily hep-ph/0309134}}].

\bibitem{Hahn-Woernle:2008tsk}
F.~Hahn-Woernle and M.~Plumacher, \emph{{Effects of reheating on
  leptogenesis}},
  \href{https://doi.org/10.1016/j.nuclphysb.2008.07.032}{\emph{Nucl. Phys. B}
  {\bfseries 806} (2009) 68} [\href{https://arxiv.org/abs/0801.3972}{{\ttfamily
  0801.3972}}].

\bibitem{Antusch:2010mv}
S.~Antusch, J.P.~Baumann, V.F.~Domcke and P.M.~Kostka, \emph{{Sneutrino Hybrid
  Inflation and Nonthermal Leptogenesis}},
  \href{https://doi.org/10.1088/1475-7516/2010/10/006}{\emph{JCAP} {\bfseries
  10} (2010) 006} [\href{https://arxiv.org/abs/1007.0708}{{\ttfamily
  1007.0708}}].

\bibitem{Croon:2019dfw}
D.~Croon, N.~Fernandez, D.~McKeen and G.~White, \emph{{Stability, reheating and
  leptogenesis}}, \href{https://doi.org/10.1007/JHEP06(2019)098}{\emph{JHEP}
  {\bfseries 06} (2019) 098}
  [\href{https://arxiv.org/abs/1903.08658}{{\ttfamily 1903.08658}}].

\bibitem{Occam:1495nn}
W.~Occam, \emph{{Quaestiones et decisiones in quattuor libros Sententiarum
  Petri Lombardi}} (1495).

\bibitem{Planck:2018vyg}
{\scshape Planck} collaboration, \emph{{Planck 2018 results. VI. Cosmological
  parameters}},
  \href{https://doi.org/10.1051/0004-6361/201833910}{\emph{Astron. Astrophys.}
  {\bfseries 641} (2020) A6}
  [\href{https://arxiv.org/abs/1807.06209}{{\ttfamily 1807.06209}}].

\bibitem{Lyth:2009zz}
D.H.~Lyth and A.R.~Liddle, \emph{{The primordial density perturbation:
  Cosmology, inflation and the origin of structure}} (2009).

\bibitem{Kawasaki:2000en}
M.~Kawasaki, K.~Kohri and N.~Sugiyama, \emph{{MeV scale reheating temperature
  and thermalization of neutrino background}},
  \href{https://doi.org/10.1103/PhysRevD.62.023506}{\emph{Phys. Rev. D}
  {\bfseries 62} (2000) 023506}
  [\href{https://arxiv.org/abs/astro-ph/0002127}{{\ttfamily
  astro-ph/0002127}}].

\bibitem{Hannestad:2004px}
S.~Hannestad, \emph{{What is the lowest possible reheating temperature?}},
  \href{https://doi.org/10.1103/PhysRevD.70.043506}{\emph{Phys. Rev. D}
  {\bfseries 70} (2004) 043506}
  [\href{https://arxiv.org/abs/astro-ph/0403291}{{\ttfamily
  astro-ph/0403291}}].

\bibitem{Esteban:2018azc}
I.~Esteban, M.C.~Gonzalez-Garcia, A.~Hernandez-Cabezudo, M.~Maltoni and
  T.~Schwetz, \emph{{Global analysis of three-flavour neutrino oscillations:
  synergies and tensions in the determination of $\theta_{23}$, $\delta_{CP}$,
  and the mass ordering}},
  \href{https://doi.org/10.1007/JHEP01(2019)106}{\emph{JHEP} {\bfseries 01}
  (2019) 106} [\href{https://arxiv.org/abs/1811.05487}{{\ttfamily
  1811.05487}}].

\bibitem{Casas:2001sr}
J.A.~Casas and A.~Ibarra, \emph{{Oscillating neutrinos and $\mu \to e,
  \gamma$}}, \href{https://doi.org/10.1016/S0550-3213(01)00475-8}{\emph{Nucl.
  Phys. B} {\bfseries 618} (2001) 171}
  [\href{https://arxiv.org/abs/hep-ph/0103065}{{\ttfamily hep-ph/0103065}}].

\bibitem{Ibarra:2003up}
A.~Ibarra and G.G.~Ross, \emph{{Neutrino phenomenology: The Case of two
  right-handed neutrinos}},
  \href{https://doi.org/10.1016/j.physletb.2004.04.037}{\emph{Phys. Lett. B}
  {\bfseries 591} (2004) 285}
  [\href{https://arxiv.org/abs/hep-ph/0312138}{{\ttfamily hep-ph/0312138}}].

\bibitem{Kolb:1990vq}
E.W.~Kolb and M.S.~Turner, \emph{{The Early Universe}}, vol.~69 (1990),
  \href{https://doi.org/10.1201/9780429492860}{10.1201/9780429492860}.

\bibitem{Davidson:2002qv}
S.~Davidson and A.~Ibarra, \emph{{A Lower bound on the right-handed neutrino
  mass from leptogenesis}},
  \href{https://doi.org/10.1016/S0370-2693(02)01735-5}{\emph{Phys. Lett. B}
  {\bfseries 535} (2002) 25}
  [\href{https://arxiv.org/abs/hep-ph/0202239}{{\ttfamily hep-ph/0202239}}].

\bibitem{Starobinsky:1980te}
A.A.~Starobinsky, \emph{{A New Type of Isotropic Cosmological Models Without
  Singularity}},
  \href{https://doi.org/10.1016/0370-2693(80)90670-X}{\emph{Phys. Lett. B}
  {\bfseries 91} (1980) 99}.

\bibitem{Kallosh:2013hoa}
R.~Kallosh and A.~Linde, \emph{{Universality Class in Conformal Inflation}},
  \href{https://doi.org/10.1088/1475-7516/2013/07/002}{\emph{JCAP} {\bfseries
  07} (2013) 002} [\href{https://arxiv.org/abs/1306.5220}{{\ttfamily
  1306.5220}}].

\bibitem{Drees:2017iod}
M.~Drees and F.~Hajkarim, \emph{{Dark Matter Production in an Early Matter
  Dominated Era}},
  \href{https://doi.org/10.1088/1475-7516/2018/02/057}{\emph{JCAP} {\bfseries
  02} (2018) 057} [\href{https://arxiv.org/abs/1711.05007}{{\ttfamily
  1711.05007}}].

\bibitem{Pilaftsis:2003gt}
A.~Pilaftsis and T.E.J.~Underwood, \emph{{Resonant leptogenesis}},
  \href{https://doi.org/10.1016/j.nuclphysb.2004.05.029}{\emph{Nucl. Phys. B}
  {\bfseries 692} (2004) 303}
  [\href{https://arxiv.org/abs/hep-ph/0309342}{{\ttfamily hep-ph/0309342}}].

\bibitem{Barger:1987nn}
V.D.~Barger and R.J.N.~Phillips, \emph{{Collider Physics}} (1987).

\end{thebibliography}\endgroup
\end{document}